\documentclass[nofootinbib,twocolumn,prd]{revtex4}
\usepackage{hyperref}
\usepackage{amsmath}
\usepackage{graphicx}
\usepackage{color}
\usepackage{ifpdf}

\newcommand\optional[1]{}
\newcommand\ForInternalReference[1]{}

\newcommand\unit[1]{\, {\rm #1}}

\newcommand\Y[1]{Y^{(#1)}}

\newcommand\avL{\left< L_{(a} L_{b)} \right>}
\newcommand\WeylScalar{{\psi_4}}
\newcommand\FourierWeylScalar{{\tilde{\psi}_4}}

\newcommand\qmstate[1]{\left|#1\right>}

\newcommand\qmstateproduct[2]{\left<#1|#2\right>}

\def\bbh#1{binary black hole#1 (BBH#1)\gdef\bbh{BBH}}
\def\bh#1{black hole#1 (BH#1)\gdef\bh{BH}}

\begin{document}
\title{Is J enough? Comparison of gravitational waves emitted along the total angular momentum direction with other
  preferred orientations}
\author{R.\ O'Shaughnessy}
\affiliation{Center for Gravitation and Cosmology, University of Wisconsin-Milwaukee,
Milwaukee, WI 53211, USA}
\email{oshaughn@gravity.phys.uwm.edu}
\author{ J. Healy}
\author{L. London}
\author{ Z. Meeks}
\author{D. Shoemaker}
\affiliation{Center for Relativistic Astrophysics,
Georgia Tech, Atlanta, GA 30332, USA}

\begin{abstract}
The gravitational wave signature emitted from a merging binary depends on the orientation of an observer
relative to the binary.    Previous studies suggest that emission along the total initial
or  total final angular momenta leads to both the strongest and simplest signal from a precessing compact binary.  
In this paper we describe a concrete counterexample: a binary with $m_1/m_2=4$, $a_1=0.6 \hat{x} = -a_2$,
placed in orbit  in the $x,y$ plane.  
We extract the gravitational wave emission along several proposed emission directions, including the initial (Newtonian) orbital
angular momentum; the final ($\simeq$ initial) total angular momentum; and
the dominant principal axis of $\avL_M$.   Using several diagnostics, we
show that the suggested preferred directions are not representative.  For example, only for a handful of other
directions ($p\lesssim  15\%$) will
the gravitational wave signal have comparable shape to the one extracted along each of these fiducial directions, as
measured by a generalized overlap ($>0.95$).   %
We conclude that the information available in just one direction (or mode) does not adequately encode the complexity of
orientation-dependent emission for even short signals from merging black hole binaries.
Future investigations of precessing, unequal-mass binaries should carefully explore and model their
orientation-dependent emission.
\end{abstract}

\maketitle

\section{Introduction}
Numerical simulations of merging compact binaries produce a gravitational wave signature that depends on the orientation
of the binary relative to the line of sight.
For nonprecessing comparable-mass binaries,  this orientation dependence can be well-modeled as purely quadrupolar
emission from   conjugate and orthogonal $l=|m|=2$ modes.  In this approximation, only one
complex-valued function is needed to characterize the gravitational wave signature from a binary:  either the
$(l,m)=(2,2)$ mode itself or the strain extracted along the binary angular momentum axis.
This approach has been broadly adopted when comparing nonspinning numerical relativity simulations to one another
\cite{2010RvMP...82.3069C} %
and to post-Newtonian
\cite{gr-nr-Jena-PNComparisons2007,nr-Goddard-merger-equalmassnonspinning-pncomparison-2007,2007PhRvD..76l4038B,gwastro-mergers-nr-HannamPNComparison2010}
and other  \cite{gw-mergers-approx-pn-nonspinning-NRCompare-AB} approximations; when building hybrid
waveforms that join systematic post-Newtonian approximations to numerical relativity \cite{2011CQGra..28m4002M,gwastro-mergers-nr-HybridReliability-Ohme2011};
when constructing phenomenological approximations to numerical relativity waveforms
\cite{gwastro-nr-Phenom-Lucia2010,gwastro-Ajith-AlignedSpinWaveforms,2010JPhCS.243a2007S}; 
and when searching for the gravitational wave signature of merging compact binaries with
interferometric detectors \cite{findchirppaper}.
For similar reasons,  the gravitational wave signature from a nonprecessing unequal-mass binary can also be approximated  by
$h_{+,\times}(t,\hat{J})$, the radiation along the total angular momentum axis $\hat{J}$.

Motivated by this approximation, some   have proposed that  radiation along ``the  total angular  momentum direction'' is a fiducial example of the
gravitational wave signature from a binary \cite{gw-astro-SpinAlignLundgren-2010}. %
In general, the total angular momentum  direction $\hat{J}$ is not conserved during the inspiral \cite{ACST} and
coalescence  \cite{gr-nr-io-review-Rezzolla2008,2010CQGra..27k4006L,2008PhRvD..78b4017B}.
To a good approximation, however, the orbital angular momentum often  precesses around the total angular momentum
\cite{ACST}.   During the inspiral, the binary  radiates predominantly along the instantaneous  orbital angular
  momentum
  \cite{WillWiseman:1996,ACST,gwastro-mergers-nr-ComovingFrameExpansionSchmidt2010,gwastro-mergers-nr-Alignment-BoyleHarald-2011},
which precesses around the total angular momentum.
 In
this case, previous direct and indirect calculations suggest that that strong, simple emission will occur along the
\emph{average} direction, which for simple precession corresponds to the total angular momentum   \cite{gw-astro-SpinAlignLundgren-2010,gwastro-mergers-nr-Alignment-ROS-Methods}.
In this paper, we provide an explicit counterexample, for which  $\hat{J}$ neither corresponds   to nor produces radiation
similar to the peak emission direction.
We contrast $\hat{J}$ with  the time-dependent and signal-averaged directions from $\avL$
\cite{gwastro-mergers-nr-Alignment-ROS-Methods}, which better reflects the peak emission direction.
 Finally, we explicitly demonstrate that higher harmonics and precession prevent a significant fraction of orientations from being well-fit as the
linear superposition of only two basis waveforms in an ``antipodal decomposition.''
The orientation-dependent emission of precessing waveforms is unavoidably more complicated than the response of their
nonprecessing counterparts.

\subsection{Context}

Merging binaries imprint their properties on the gravitational waves radiated to infinity in all directions \cite{2010RvMP...82.3069C}.  
In practice, however,  gravitational wave detectors have access to only one line of sight, sometimes only one
polarization.   Under the right circumstances, two physically distinct configurations can produce very similar emission in the detectors' sensitive band.  These
near-degeneracies complicate source parameter estimation  
\cite{2008CQGra..25r4011V,2009CQGra..26k4007R,2007PhRvD..75f2004R,2011PhRvD..84b2002L,gwastro-mergers-PE-Aylott-LIGOATest,2011ApJ...739...99N}
and thus the interpretation of gravitational wave measurements for astrophysics.

Conversely, a single source can produce highly dissimilar signals along different lines of sight.  
By contrast, the signal template used in real searches for nonprecessing binaries in gravitational wave data
 assumes a universal  form for all orientations: a linear superposition of  two basis functions, the ``cosine'' and
``sine'' chirps \cite{findchirppaper}.    For nonprecessing binaries, this signal model is physically well-motivated [see the Appendix]; the
relative weights of the modes are in one-to-one relation to the source inclination relative to the line of sight.
In this paper we use a concrete example where this decomposition fails.  For the example described in this paper, more than two functional degrees of freedom are
needed to reproduce the gravitational wave signal accurately for all lines of sight.

\section{Orientation-dependent emission}

\begin{figure}
\includegraphics[width=\columnwidth]{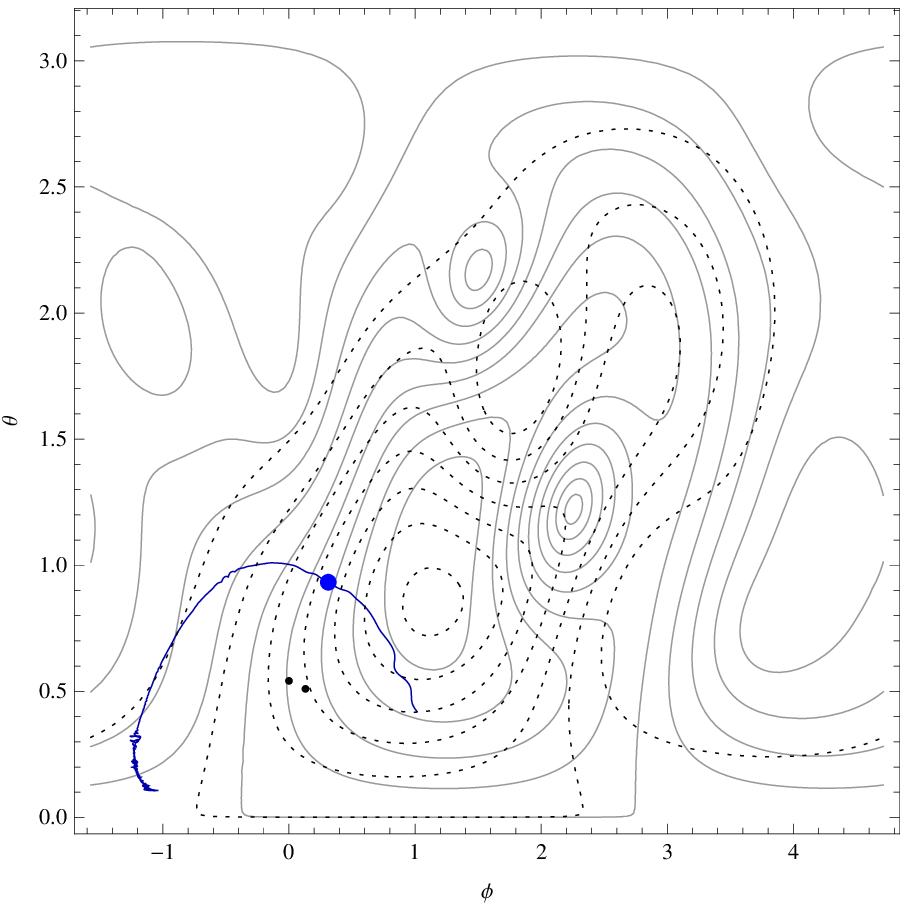}
\includegraphics[width=\columnwidth]{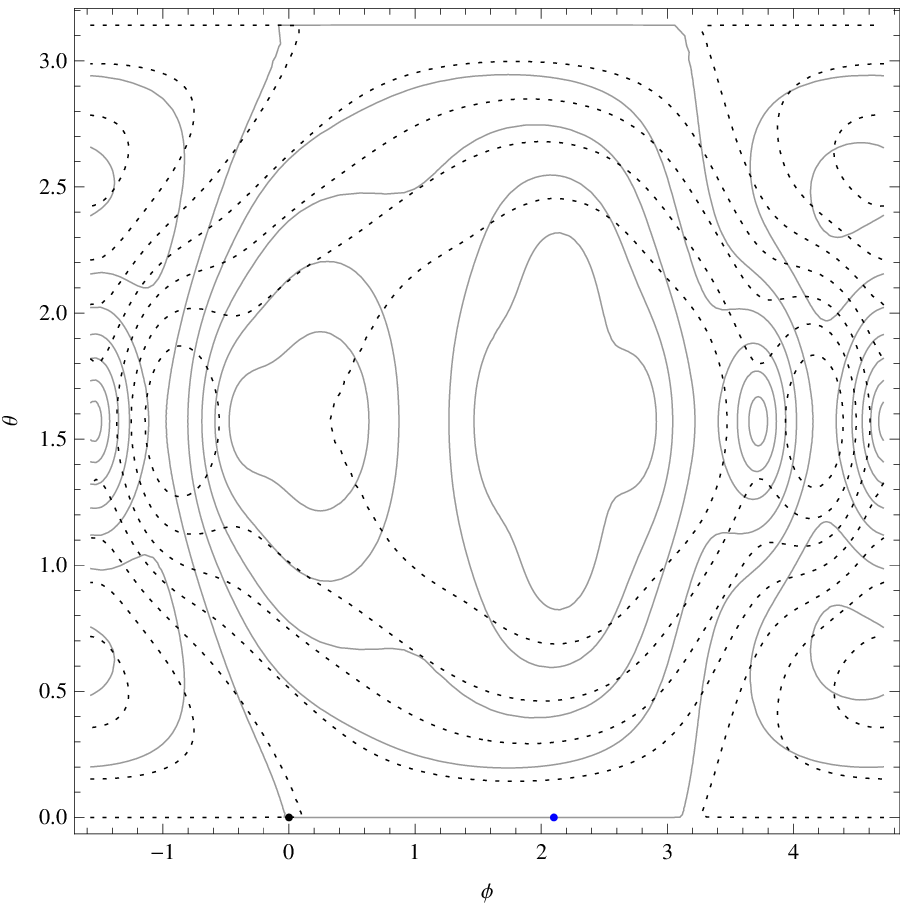}
\caption{\label{fig:Amplitudes:Time}\textbf{Time-domain contours} of $|\psi_4|$ (solid) and $dE/dtd\Omega$ (dashed), evaluated at the peak of the
  $(l,m)=(2,2)$ mode.  Both quantities are normalized to $1$ at $\theta=\phi=0$.  Solid black dots indicate the location
  of the initial and final total angular momenta.  The solid blue dot indicates the preferred axis identified by
  $\avL_t$ at this time.  The dark blue path indicates how this preferred orientation evolves with time, several hundred
  $M$ before and $30M$ after this reference time.
 The \emph{top panel} corresponds to the fiducial spinning binary.  The bottom panel shows a binary with comparable
  mass ratio ($m_1/m_2=4$) but without any spin.
}
\end{figure}

\begin{figure}
\includegraphics[width=\columnwidth]{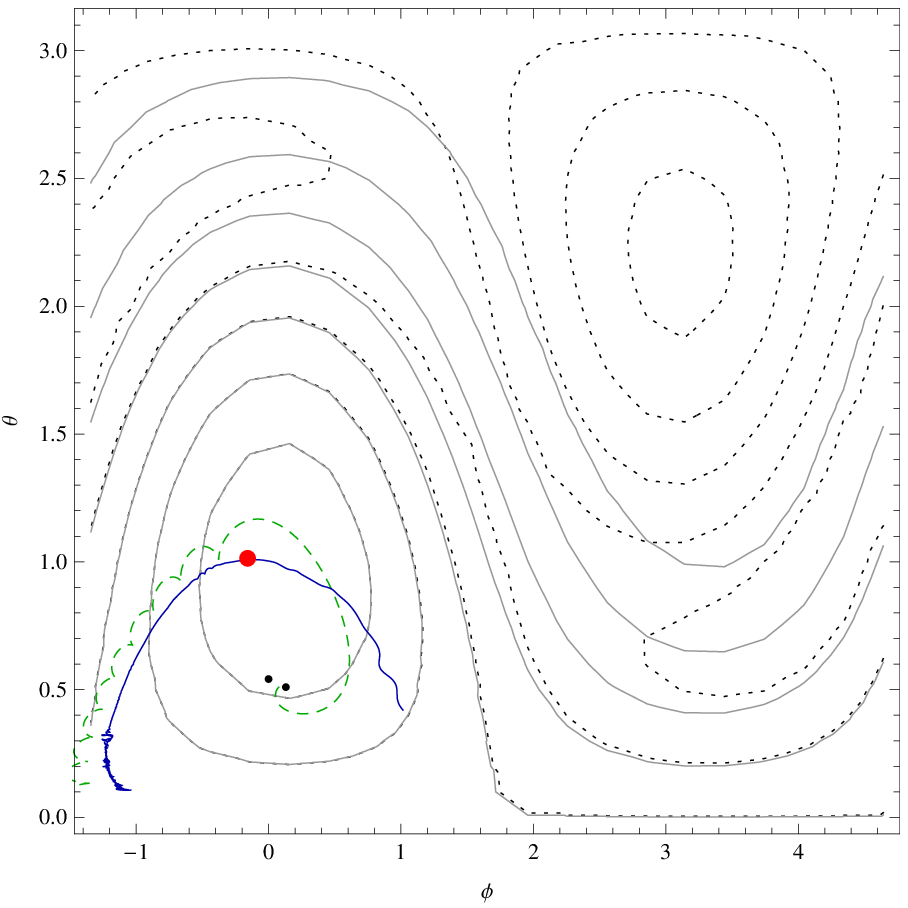}
\includegraphics[width=\columnwidth]{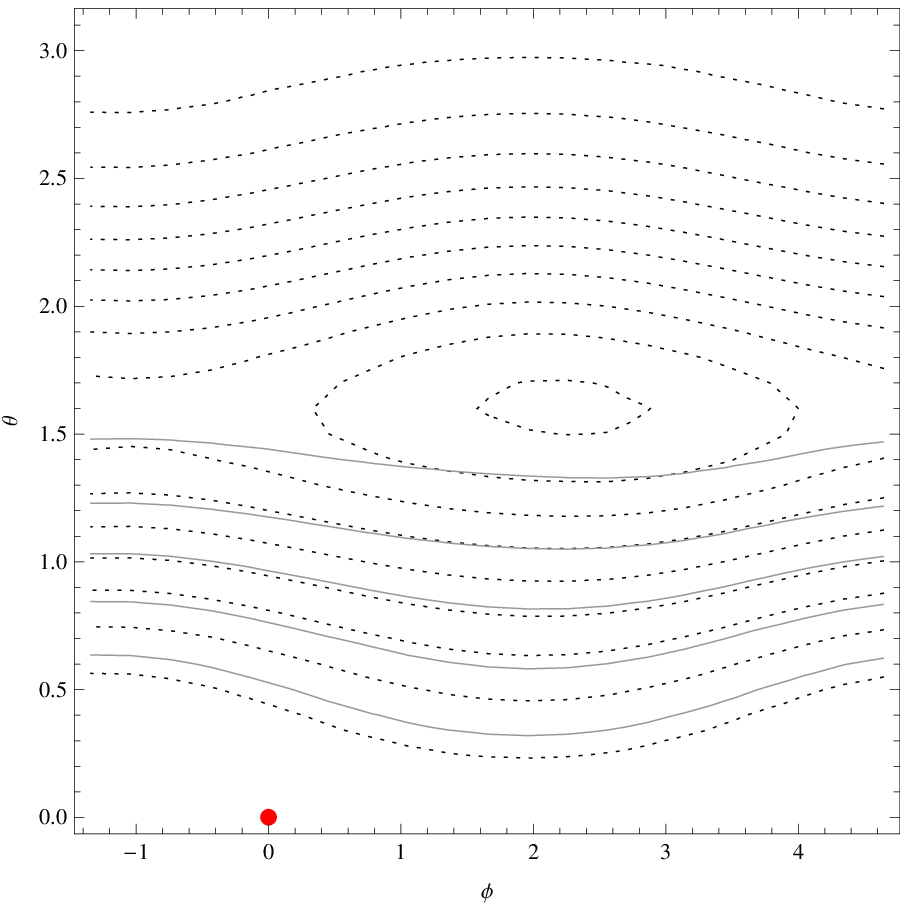}
\caption{\label{fig:Amplitudes:Frequency}\textbf{Detector-weighted contours} of $\rho$ (dotted) and $\rho_+$ (solid),
  evaluated at $M=100 M_\odot$,  plotted versus the simulation's asymptotic spherical polar coordinates (i.e., aligned with the
  initial orbital angular momentum).   Solid black dots indicate the location
  of the initial and final total angular momenta.  The solid red dot indicates the preferred axis identified by
  $\avL_{M,+}$ [Eq. (\ref{eq:avL:Plus})] at this time; nearly identical preferred axes are identifies by $\avL_M$ and
  $\avL_{M-}$.
 As in Fig. \ref{fig:Amplitudes:Time}, the solid blue path indicates the evolution of $\avL$ versus time.  The
  dashed green path indicates the orientation of the (Newtonian, coordinate) orbital angular momentum $\propto {\bf
    r}\times {\bf v}$.
The \emph{top panel} corresponds to the fiducial spinning binary, while the \emph{bottom panel} shows an nonspinning
binary with comparable mass ratio ($q=4$).
 In the bottom panel,  the preferred orientation identified by $\avL_{M,+}$ is slightly above  the $\theta=0$ line;
  the two directions identified by $\avL_{M,\pm}$ are not precisely antipodal.
}
\end{figure}

We simulated the evolution of a \bbh{}, with total mass  $M = m_1 + m_2 = 1.0$ and mass ratio $m_1/m_2=4$, initiated on the $x$-axis, 
with initial separation $d=9$ and initial dimensionless spin vectors ${\bf a}_1 = 0.6 \hat{x}=-{\bf a}_2$.  For
reference, the initial and final total angular momenta are
\begin{eqnarray}
J_i &=& 0.36 \hat{x}+0.60 \hat{z}\\
J_f &=& 0.27 \hat{x} + 0.036 \hat{y} + 0.49 \hat{z}
\end{eqnarray}
Initial data was evolved with  \texttt{Maya}, which was used in previous \bbh{} studies \cite{2007CQGra..24...33H,Herrmann:2007ex,Herrmann:2007ac,Hinder:2007qu,Healy:2008js,Hinder:2008kv,Healy:2009zm,Healy:2009ir,Bode:2009mt}.
The grid structure for each run consisted of  10 levels of refinement 
provided by \texttt{CARPET} \cite{Schnetter-etal-03b}, a 
mesh refinement package for \texttt{CACTUS} \cite{cactus-web}. 
Sixth-order spatial finite differencing was used with the BSSN equations 
implemented with Kranc \cite{Husa:2004ip}. The outer boundaries are 
located at 307M. Each simulation was performed with a resolution of $M/140$
on the finest refinement level, with each successive level's resolution
decreased by a factor of 2.  
In the text, we present results for the fiducial $r=90$ extraction radius and for this resolution.   The
comparatively large  differences we describe exist at all large extraction radii and simulation resolutions we have explored.
Similarly, though we adopt fiducial snapshot parameters (time and mass) to generate the contours in Figures
\ref{fig:Amplitudes:Time} and \ref{fig:Amplitudes:Frequency}, we find similar behavior at neighboring masses and times.

This simulation has markedly anisotropic emission.    Figure \ref{fig:Amplitudes:Time} shows contours of $|\psi_4|$ (solid) and $dE/dtd\Omega \propto |\int
\psi_4 dt|^2$ (dashed), superimposed with black dots indicating the total angular momentum direction and a blue dot
indicating the dominant eigenvector of $\avL$  \cite{gwastro-mergers-nr-Alignment-ROS-Methods}.  This tensor is defined
by
\begin{eqnarray}
\label{eq:avL}
\avL_{M} &=& 
 \frac{\int_{-\infty}^{\infty} df \int d\Omega \FourierWeylScalar^*(f) L_{(a}L_{b)} \FourierWeylScalar(f)/S_h
  }{
   \int_{-\infty}^\infty df \int d\Omega  |\FourierWeylScalar(f)|^2/S_h
}
\end{eqnarray}
where $S_h$ is the noise power spectrum of a candidate interferometer.
None of these three directions corresponds to the  directions of peak transverse tidal force or
radiated power. 

 For clarity we only show contours for a single time.  Over the inspiral, the beampattern evolves slightly, with the
  orientation of strongest $\psi_4$  and $dE/dt$ roughly following the path of our preferred orientation (blue line).
 For all times simulated, the largest $\psi_4$ and $dE/dtd\Omega$ is never along the total angular momentum.

For comparison, in the bottom panel of Figure \ref{fig:Amplitudes:Time} we show contours for a simulation with
comparable mass ratio in the absence of spin.   Though vertically symmetric, the functional form of $\psi_4$ 
and $dE/dtd\Omega$ is largest at specific angle $\phi$ that evolves along with  the (retarded) separation vector.     %
For this particular timeslice,  the peak occurs at $(\theta,\phi) \simeq (0.5,1.5)$, significantly offset from the angular momentum axis.
This shape rotates at the orbital period through merger.  By
contrast, all three preferred orientations ($L,J$ and $\hat{V}$, the principal eigendirection of $\avL$) always lie
along the $\hat{z}$ axis.  
For data analysis purposes, a more relevant orientation-dependent quantity is the coherent signal-to-noise accessible to
a
particular detector network along that line of sight.  Under the simplifying assumptions that two identical
interferometers misaligned by $\pi/4$ intercept gravitational wave
emission perpendicular to the line of sight, an idealized orientation-dependent signal-to-noise ratio $\rho(\hat{n})$ is
given by \cite{gwastro-spins-rangefit2010,gwastro-mergers-nr-Alignment-ROS-Methods}\footnote{For comparison, the
  orientation-averaged signal to noise $\bar{\rho}$ described in \citet{gwastro-spins-rangefit2010} is precisely the
 orientation  average $\left<\rho^2\right>$  of our signal-to-noise ratio over all emission directions.}
\begin{eqnarray}
\rho^2&\equiv& 2\int_{-\infty}^{\infty} \frac{df}{S_h} \frac{|\psi_4(f,\hat{n})|^2}{(2\pi f)^4}
\end{eqnarray}
where $S_h$ is the noise power spectrum of each of the two interferometers.
 (This expression is simply the  sum of the signal to noise due to each detector ($\rho_1^2+\rho_2^2$), using
$\WeylScalar=\partial_t^2(h_+-ih_\times)$ to relate the sum to a quantity naturally produced by simulations.)
The dotted contours in Figure \ref{fig:Amplitudes:Frequency} show contours of constant $\rho$, 
adopting a simple analytic model for the initial LIGO noise curve \cite{gw-astro-mergers-NRParameterEstimation-Nonspinning-Ajith} and a fiducial total mass
$M=100 M_\odot$ for the binary to scale our (dimenionless) simulations.
Again, neither the total nor (Newtonian) orbital angular momentum direction seems significant with respect to the detectable signal
power.

The contours of $\rho^2$ and its peak location inevitably are qualitatively similar to contours of $dE/dfd\Omega$, the energy
radiated per unit frequency.  The expression for $\rho^2(M)$ can be rewritten as a
suitable frequency-weighted average of $dE/dfd\Omega  \propto
|\FourierWeylScalar/f |^2$.  For a suitable choice of $f$,
contours in the neighborhood of the maximum are qualitatively similar.    

Two- or more detector networks are generally sensitive to gravitational wave polarization.  Our cartoon two-detector
network has equal sensitivity to both polarizations.  Such a network can automatically project out left- or right-handed
signal.
Motivated by sensitivity to helicity, we define two signal to noise ratios $\rho_{\pm}$ that measure sensitivity to
right and left handed emisison, respectively:
\begin{eqnarray}
\rho_+^2&\equiv& 2\int_{0}^{\infty} \frac{df}{S_h} \frac{|\psi_4(f,\hat{n})|^2}{(2\pi f)^4}\\
\rho_-^2&\equiv& 2\int_{-\infty}^{0} \frac{df}{S_h} \frac{|\psi_4(f,\hat{n})|^2}{(2\pi f)^4}
\end{eqnarray}
At each instant, right-handed emission occurs along the orbital angular momentum direction, left-handed
opposite.  These expressions encode the contributions of right-handed and left-handed emission to the
total SNR $\rho^2=\rho_-^2+\rho_+^2$.
The solid curves in Figure \ref{fig:Amplitudes:Frequency} are contours of $\rho_+$.  Their shape and similarity to
contours of $\rho$ demonstrate that each of the two local extrema of
$\rho^2$ is produced near-exclusively  by emission with a single handedness.
Neither local extrema corresponds to the direction of $J$ or the (Newtonian) orbital angular momentum.
By contrast, the nearly-antipodal extrema do correspond to the orientation of the principal eigenvalue of $\avL_M$
\cite{gwastro-mergers-nr-Alignment-ROS-Methods}, shown as a red dot in Figure 
 \ref{fig:Amplitudes:Frequency}.

To summarize, for our fiducial binary, we have examined the gravitational wave signature along  three common choices for
the preferred orientation: the principal axis of $\avL_M$ as well as the (Newtonian) orbital and total angular momentum directions.
Neither angular momentum direction corresponds to a direction extremizing  strain, energy flux, or detectable signal to
noise.
By contrast, the direction identified by $\avL_M$  nearly corresponds to  a local maximum of  the signal-to-noise  $\rho^2$.
We anticipate and will show below that the direction identified by $\avL_M$ is ``more representative'' of the
orientation-dependent emission, in that a greater fraction of the orientation-dependent emission can be well-fit by it.
However, none of the three directions can adequately fit all the complexities of this binary's orientation-dependent
emission, even for one representative mass $M=100 M_\odot$.

\section{Contrasting different orientations }
\label{sec:Orient}
Most  variations in the gravitational wave signature or amplitude along different emission directions
are too subtle to be distinguished by gravitational wave networks at the amplitude of astrophysically plausible signals.
Generally, distinguishability of signals depends strongly on the detector network, signal model, and source location
relative to the earth.  %
Rather than address this complicated issue in general, we adopt a simple physically-motivated proof of concept
calculation.  
The complications associated with real detector networks will, if anything, further break any symmetries connecting the
natural emission direction to  $\hat{J}$.

\begin{figure}
\includegraphics[width=\columnwidth]{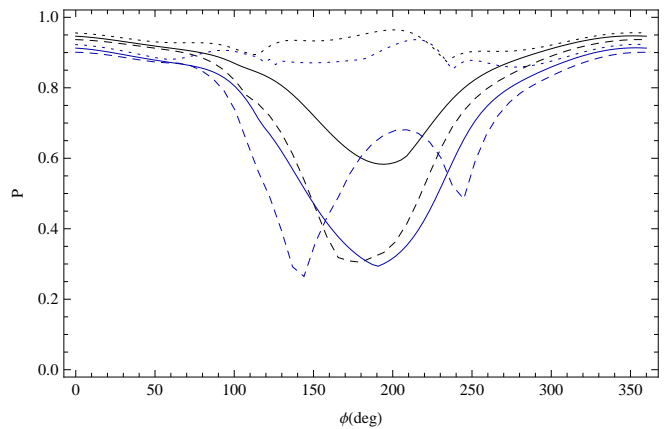}
\caption{\label{fig:ComplexIsSane}\textbf{Comparison of overlap for complex versus standard match}: These curves compare
  three different criteria to distinguish between emission along $\hat{z}$ and $\hat{n}(\theta,\phi)$, shown for
  $\theta=40.5\unit{deg}$ (black) and $53\unit{deg}$ (blue).  The solid curve is the normalized complex overlap
  $|P(\hat{z},\hat{n})|$.  The dotted curve and dashed curves are the ``typical'' and ``minimax'' match constructed from $h_+,h_\times$ along the two lines
  of sight \cite{2007PhRvD..76h4020V}.
}
\end{figure}

First and foremost, we introduce a complex overlap and its natural maximization in time and \emph{polarization}, a tool
which efficiently eliminates two extrinsic parameters of generic signals even in the presence of higher harmonics.
Though we adopt this complex overlap for technical convenience and its explicit relationship with $\avL$, all results
described below also follow from the standard  inner product  on a single polarization at a time
[Eq. (\ref{eq:StanardInnerProduct}); see 
\cite{CutlerFlanagan:1994,2007PhRvD..76h4020V}].
Technically,  subtle differences arise due to the number of data streams and the order of maximization over extrinsic parameters: time and polarization
(two-detector complex) or time and phase (single-detector match).
Practically speaking, however, the two expressions often return comparable results, particularly when one of the two
signals is nearly circularly polarized as will be the case below.  As an example, in Figure \ref{fig:ComplexIsSane} we
compare the complex match $P$ to the ``typical'' and ``minimax'' matches
\cite{2007PhRvD..76h4020V}.
In regions of high complex overlap, the two are qualitatively similar.  In regions with low complex overlap, the ``typical'' and
``minimax'' matches differ substantially, indicating the two linear polarizations along that line of sight are distinguishable and
comparable in magnitude.

\subsection{Complex overlap}

For each emission direction $\hat{n}$,  we once again assume two identical interferometers are placed, oriented by $\pi/4$ relative to one another and
perpendicular to the incident signal.  We coherently compare the (noise-free) signal expected along any pair of orientations
with a complex inner product motivated by the detector's noise power spectrum \cite{gwastro-mergers-nr-Alignment-ROS-Methods}.

For our purposes, numerical relativity simulations take as inputs binary black hole parameters and desired line of sight
(denoted by $\lambda$) and return  the Weyl scalar $\psi_4(t)$,  a complex-valued function of time evaluated along that
line of sight.
For any pair of simulations and lines of sight,  we compare $\WeylScalar$ and $\WeylScalar'$ by a complex
overlap
\begin{subequations}
\label{eq:def:Overlap}
\begin{eqnarray}
P(\lambda,\lambda')  &\equiv & 
 \frac{\left( r \WeylScalar |r \WeylScalar' \right) }{|r\WeylScalar| |r\WeylScalar|} \\
(A,B) &\equiv&  \int_{-\infty}^{\infty} 2 \frac{df}{(2\pi f)^4 S_h} \tilde{A}(f)^* \tilde{B}(f)
\end{eqnarray}
\end{subequations}
where $S_h$ is a detector strain noise power spectrum.  
 In this and subsequent expressions we used unprimed and primed variables to distinguish between the two waveforms being
compared, involving potentially distinct parameters $\lambda'$ and lines of sight $\hat{n}'$.
For simplicity and to avoid ambiguity, in this paper, we adopt
a semianalytic model for the initial LIGO 
sensitivity \cite{gw-astro-mergers-NRParameterEstimation-Nonspinning-Ajith}.
As with the single-detector overlap, the complex overlap can be  maximized over the event time and \emph{polarization}  ($t_c,\psi_c)$ and  by a simple fourier transform:
\begin{eqnarray}
P_{max}&\equiv& \text{max}_{t_c,\psi_c} |P| \\
& =&\frac{1}{|\WeylScalar| |\WeylScalar'|} 
\left| \int 2 \frac{df}{(2\pi f)^4 S_h} \FourierWeylScalar(f)^* \FourierWeylScalar(f) e^{i(2\pi f t_c + \psi_c)} \right|  \nonumber
\end{eqnarray}

For a network with isotropic sensitivity to both gravitational wave polarizations,\footnote{Currently-planned
  gravitational-wave detector networks are primarily sensitive to one polarization for most sky
locations \cite{gwic-roadmap,2011CQGra..28l5023S}.  Future
  gravitational wave networks will try to achieve comparable sensitivity in both polarizations, to better measure source
inclination and distance \cite{gwic-roadmap}. } this complex-valued expression
naturally characterizes the ability of the network to differentiate signals.  
The real part of this expression corresponds to  a coherent sum of the conventional overlaps of the two gravitational wave
polarizations generated from $\psi$:
\begin{eqnarray}
\label{eq:ComplexHasMoreInformationThanStandard}
\text{Re}(\WeylScalar,\WeylScalar') & =& \qmstateproduct{h_+}{h_+'} + \qmstateproduct{h_\times}{h_{\times}'} \\
\tilde{h}_+ + i \tilde{h}_\times &=& - \psi(f)/(2\pi f)^2 \\
\label{eq:StanardInnerProduct}
\qmstateproduct{A}{B} &=& 2\int_{-\infty} ^{\infty} df \tilde{A}^* \tilde{B}/S_h
\end{eqnarray}
where $\qmstateproduct{A}{B}$ is the standard single-detector overlap of gravitational wave strain. 
This expression also relies on the gauge-invariant output of
gravitational wave simulations, without any complications associated with conversion  $\WeylScalar\rightarrow h(t)$.
Critically, this expression naturally respects rotational symmetry around the line of sight: if the two simulations are
rotated by $\gamma,\gamma'$, then $\psi_4\rightarrow e^{-2i\gamma}\psi_4$ and similarly, so $P$ changes by a pure phase:
\[
P(R\lambda,R'\lambda') = e^{-2i(\gamma'-\gamma)}P(\lambda,\lambda')
\]
Equivalently, the absolute value $|P|$ corresponds to choosing the (physically arbitrary) orientations of the two
simulations such that $P$ is real.
Similarly, the complex overlap interacts  transparently with a decomposition of the NR signal into (complex) spin-weighted spherical
harmonics.   For example, as with the conventional single-detector signal, a network with equal sensitivities to two
polarizations has a network Fisher matrix along the line of sight proportional to
\begin{eqnarray}
\Gamma_{ab}=\left(\partial_a \WeylScalar,\partial_b \WeylScalar \right)
\end{eqnarray}
For rotations,  the Fisher matrix can be reorganized as an action of
SU(2) generators on $\WeylScalar$  \cite{gwastro-mergers-nr-Alignment-ROS-Methods}:
\begin{eqnarray}
\label{eq:Fisher:Angles}
 \Gamma_{ab} =  \left(L_a \WeylScalar, L_b \WeylScalar \right)
\end{eqnarray}
This action can be evaluated algebraically using the known properties of SU(2).

Qualitatively speaking, the complex overlap is considerably more discriminating than a single-polarization overlap.  As
an example, because the complex overlap naturally distinguishes between helicities, the complex overlap between the
$\hat{z}$ and $-\hat{z}$ directions of an inspiralling nonprecessing binary is nearly 0.  By contrast, a single-detector overlap
would be nearly unity, as one polarization cannot distinguish between left- and right-handed signals.
Finally, in the complex overlap, maximization over the physically irrelevant polarization angle $\gamma$ corresponds to
an absolute value.\footnote{For a single-detector overlap,  maximization over polarization
   requires  diagonalizing a quadratic form for basis waveforms  \cite{2007PhRvD..76h4020V}.  Physically, this process
   corresponds to choosing a detector orientation aligned with the natural principal axes of the waveform, projected on
   the plane of the sky.     When multiple harmonics are present,
   maximization of single-detector overlaps becomes an
  arduous process; see, e.g., the Appendix of \citet{gwastro-nr-SemiAnalyticModel-McWilliams2010}.  Finally,
  for the single-detector overlap, efficient maximization over polarization is generally incompatible with fast
  (fourier-transform-based) determination of the coalescence time \cite{2007PhRvD..76h4020V}. 
}

\subsection{Area of high overlap}
For each of the three reference orientations identified above  -- $\hat{v}=\hat{z}$, the initial (Newtonian) orbital angular momentum
direction; $\hat{J}$, the nearly-conserved final angular momentum direction; and $\hat{V}$, the preferred orientation
identified by $\avL_{M+}$ --  we compute
$P_{max}(\hat{v},\hat{n})$ everywhere on the sphere.

\begin{figure}
\includegraphics[width=\columnwidth]{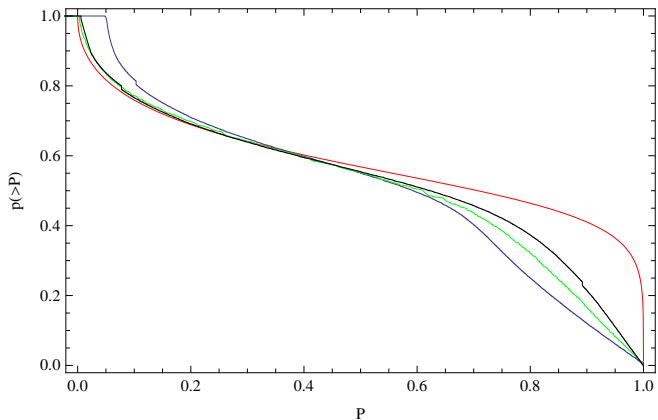}
\caption{\label{fig:AreaVersusOverlap}\textbf{Area of high overlap}: Fraction $p$ of the sphere for which
 the overlap  $P_{max}(\hat{v},\hat{n})$ is above a given threshold $P$, versus the threshold, for $\hat{v}=\hat{z}$
 (blue), $\hat{J}$ (green), and the preferred axis of $\avL_M$ (black).
For comparison, the thin red curve shows the corresponding result for a binary radiating only  conjugate and orthogonal $(2,2)$ and
$(2,-2)$ modes and a reference axis $\hat{v}=\hat{z}$; see the appendix for a derivation.  For this idealized nonprecessing binary, more than a third of all
orientations have better than $P=0.95$ (complex) match with the
$(2,2)$ mode; all are well-fit with an antipodal decomposition.   For the three reference orientations used here,
the gravitational wave signal in that direction closely resembles the gravitational wave signal for only a small
neighborhood surrounding that point.
Results for simulations are estimated using $\simeq 10^4$  %
 candidate viewing directions, each of which includes all $l\le
4$ contributions to $\WeylScalar(\hat{n})$.
}
\end{figure}

For comparison, as indicated by the red curve in Figure \ref{fig:AreaVersusOverlap}, for a nonprecessing binary the emission along $+\hat{J}=\hat{z}=\hat{V}$ alone can fit
about one third of all emission orientations, weighted equally in area; and about 48\%, weighted uniformly in $\rho^3$
(i.e., by equal
probability of detection in the local universe).
Summing over both antipodal directions, almost all detected emission from a nonprecessing binary is well-fit by either of
$\WeylScalar(\pm \hat{J})$. 

By contrast, Figure \ref{fig:AreaVersusOverlap} shows that each of these directions provides a good fit ($P>0.95$) only for a small
area surrounding each reference point.    
 The gravitational wave signal along the (Newtonian) orbital and total angular momentum directions ($\hat{z},\hat{J}$)
are particularly nonrepresentative, reproducing signals emitted only in immediately neighboring directions.

\subsection{Failure of antipodal reconstruction }
Existing gravitational wave searches for nonprecessing binaries, however, do not merely fit each data set against two 
candidate signals per simulation.  Motivated by the simple functional form of mass-quadrupole-dominated emission, 
real searches \emph{project the data onto the two-dimensional subspace spanned by those two candidates}.\footnote{Their real and imaginary parts
  are known as the ``cosine'' and ``sine'' chirps.  For sufficiently low mass, these two modes  are nearly orthogonal.  Their orthogonality can be exploited to efficiently maximize in time
  and ``coalescence phase'' \cite{findchirppaper}.   At high mass too few cycles exist to
  guarantee orthogonality of the $(2,2)$ and $(2,-2)$ modes in general.}
When the candidate signals are associated with two different antipodal emission  directions ($\pm\hat{v}$)
we will denote this procedure an \emph{antipodal decomposition}.   

Algebraically, an antipodal decomposition estimates the Weyl scalar along any direction via a projection to the subspace
spanned by the two antipodes:
\begin{eqnarray}
\qmstate{N} &\equiv& \psi_4(\hat{v})/|\psi_4(\hat{v})| \\
\qmstate{S} &\equiv& \psi_4(-\hat{v})/|\psi_4(-\hat{v})| \\
\begin{bmatrix}
\rho_N \\ \rho_S 
\end{bmatrix}
&=& 
\begin{bmatrix}
\left(N,\WeylScalar(\hat{n}) \right)
\\
\left(S,\WeylScalar(\hat{n}) \right)
\end{bmatrix} \\
\WeylScalar_{est}&\equiv& {\cal P} \WeylScalar 
\nonumber \\  &&
\frac{
\left[ (\rho_N - (S,N) \rho_S) \qmstate{N} + (\rho_S - (N,S)
  \rho_N) \qmstate{S} \right] }{1-|\left(N,S)\right|^2} 
 \\
\rho_{rem}^2  &\equiv & |\WeylScalar - \WeylScalar_{est}|^2 = (\WeylScalar,(1-{\cal P})\WeylScalar)
\nonumber \\ &=&
  \rho^2 - \frac{|\rho_N|^2 + |\rho_S|^2 - 2 \text{Re} (N,S)\rho_N\rho_S^*}{1-|(N,S)|^2}
\end{eqnarray}
Practically, an antipodal projection of this kind is useful if the relative loss of SNR is small (i.e., $\rho_{rem}/\rho < 0.05$).

For nonprecessing signals, the success of an antipodal approximation along nearly \emph{any} axis is guaranteed by
parameter counting. 
Comparable-mass nonprecessing binaries are dominated by just two modes (the $(l,m)=(2,2)$ and $(2,-2)$ modes), with two coefficient
functions $\WeylScalar_{2\pm 2}$.
As any antipodal basis  are in their span, an antipodal decomposition just corresponds to a change of basis.
For nonprecessing binaries, however, the choice $\hat{v}=\hat{J}$ produces  added benefits: the antipodes are local
extrema of $|\WeylScalar|$, $dE/dtd\Omega$, and $\rho^2$.    In this frame,  trivially the angular dependence of
$\rho_{N,S}(\hat{n})$ decomposes  into roughly two spin-weighted harmonics, to the extent that higher harmonics can be neglected.
Finally, as noted previously this antipodal decomposition is \emph{physical}: the ratio $\rho_N/\rho_S$ is in one-to-one
relation with the inclination.

By constrast, for precessing binaries the signal space  generally has at least $5(=2\times2+1)$ basis vectors and thus
has $5$ complex (10 real) dimensions
\cite{BCV:PTF}.  The precession of \emph{instantaneous} $(2,2)$ and $(2,-2)$ modes aligned with $L$ projects into other $l=2$ modes in
the global frame.
As a result, the ``remaining signal power'' $\rho_{rem}$ left after an antipodal decomposition is generally nonzero and
often significant, if either precession occurs or if higher harmonics contribute significantly to overall radiation.
 Nonprecessing  binaries also emit a significant fraction of their total (band-limited) signal $\rho^2$ in higher
harmonics,  at high mass  \cite{gwastro-spins-rangefit2010} or  mass ratio \cite{gwastro-nr-SemiAnalyticModel-McWilliams2010,gwastro-mergers-nr-HannamPNComparison2010}.
However, comparable-mass nonprecessing binaries are often still well-approximated as antipodal: for most masses, they emit most strongly in two
complex-conjugate $l=|m|$ modes, with $|m|<l$ suppressed.  By contrast, precession ensures that all five $l=2$ modes are
strongly excited, for all masses.

For this simulation and reference mass ($100 M_\odot$) in particular, antipodal decompositions always lose
SNR for a significant fraction of orientations.  One way to see this is by example, decomposing the waveform along
$\hat{z},\hat{J},\hat{V}$ in an antipodal decomposition, using each of the other combinations as a basis.  %
None of these directions provides an adequate antipodal basis for reconstructing the waveform along the others.
More abstractly, antipodal decompositions fail when, even adopting a well-chosen frame (primed, below) to define the mode basis, a significant
fraction of the  orientation-averaged signal power   \cite{gwastro-spins-rangefit2010}
\begin{eqnarray}
\bar{\rho}^2 &\equiv& \int \left(\WeylScalar,\WeylScalar \right) \frac{d\Omega}{4\pi} \nonumber \\ 
&=& \sum_{lm} \frac{1}{4\pi}  \left(\WeylScalar_{lm},\WeylScalar_{lm} \right) \equiv \sum_{lm} \rho^2_{lm}
\end{eqnarray}
comes from  harmonics other than the $(2,2)'$ and $(2,-2)'$ modes.
For this simulation and reference mass, the orientation-averaged SNR $\bar{\rho}$ for a source at $100 \unit{Mpc}$ is
$\simeq 25$.  Using a (primed) frame aligned with $\hat{V}$, at least nine modes contribute significantly ($>0.5\%$,
summing to $\simeq 99\%$) to $\bar{\rho}^2$; see Table \ref{tab:Amplitudes}.  
In other words, for this simulation and reference mass, other harmonics invariably contribute a significant fraction of
the signal power, for many significant lines of sight.

\begin{table}
\begin{tabular}{lr|l|l}
 l  & m & $(\rho_{lm}/\bar{\rho})^2$  & $(\rho_{lm}/\bar{\rho})^2$ \\
    &     & precessing  & nonprecessing \\
\hline
 2 & 2 & 0.43   & 0.44 \\
 2 & -2 & 0.39  & 0.44\\
 2 & 1 & 0.067  & 0.01\\
 2 & -1 & 0.010  &0.01 \\
 2 &  0 & 0.009  & $10^{-4}$  \\
 3 & 3 & 0.039  & 0.045 \\
 3 & -3 & 0.037  &0.045 \\
 4 & 4 & 0.005  & 0.007 \\
 4 & -4 & 0.005 & 0.007
\end{tabular}
\caption{\label{tab:Amplitudes}\textbf{Multiple harmonics contribute significantly} to the orientation-averaged signal power, for our
  reference simulation and mass ($100 M_\odot$).  This table refers to mode amplitudes in a frame aligned with the prinicipal axis of
  $\avL_M$.    The second and third columns show results for our two fiducial simulations, both with $q=4$ but one
    precessing and one without any spin.    At this mass ratio, both nonprecessing and precessing binaries have
    observationally significant power in higher modes;
  see, e.g., Table V in \citet{gwastro-mergers-nr-HannamPNComparison2010}. 
}
\end{table}

We have chosen a relatively low reference mass ($100 M_\odot$), for which higher-order multiples $l>2$ produce a relatively small fraction of the
orientation-averaged  SNR ($\bar{\rho}$).  At higher masses, higher-order multipoles will produce even larger fractions of
the total power \cite{gwastro-spins-rangefit2010} and can lead to greater breakdowns of antipodal approximations.

To summarize, many simplifications extensively employed for nonprecessing binaries break down completely for generic
mergers.  In general one cannot usefully approximate the line-of-sight gravitational waveform with just two basis functions. 
As illustrated Table \ref{tab:Amplitudes} for this concrete example, the $|m|=2$ subspace is often but generally  need not be the most significant feature.
For a nonprecessing binary dominated by the $(2,2)$ and $(2,-2)$ modes, the  real and imaginary parts of the waveform
along the $z$ axis are proportional to the real and imaginary parts of $e^{2i\Phi_o}$ and therefore are related by a simple
phase shift:  the signal ($\WeylScalar(\hat{z},\Phi_o)$) produced by a binary with initial orbital phase $\Phi_o$ is a phase
shift by $\pi/4$ ($-i\WeylScalar(\hat{z},\Phi_o+\pi/4)$) of the signal produced by a binary starting with a similarly shifted binary phase. 
By contrast, for a generic binary, %
multiple harmonics contribute significantly; for nearly no line of sight does the simple phase-shift relationship
described above hold to the accuracy needed for gravitational wave data analysis.
 One
cannot maximize over the orbital reference phase $\Phi_o$ using an antipodal projection to extract coefficients of the ``cosine and sine
chirps''.   The relevant maximum over coaelscence phase can be found only if all harmonics are included \cite{gwastro-nr-SemiAnalyticModel-McWilliams2010}.

\subsection{Constrained projections onto all harmonics}
More broadly, the signal from precessing binaries can be projected onto a higher-dimensional basis.    As only four
degrees of freedom are needed to specify the line of sight (two euler angles, polarization, and distance to the source), the use of $N>4/2$  basis
vectors requires $N-2$ complex ($2(N-2)$ real) constraints, to insure the reconstructed signal estimate is consistent with what could generate
it.   

Broadly speaking, constrained maximization methods have been extensively explored in gravitational wave data analysis.
Constrained maximization methods are primarily employed in the context of interpolating a signal manifold using a basis
\cite{2010PhRvD..82d4025C,gw-astro-mergers-interpolation-FieldEtAl2011}.
One particularly relevant example is \citet{BCV:PTF}, who describe a strategy for detecting generic precessing low-mass binaries that, for each binary,
finds the best-possible line of sight from the binary by first projecting data onto $5$ basis signals, then solving
constraint equations. %
In a similar fashion, the single orientation of a particular simulation that best fits a candidate signal can be
recovered by projecting onto a simulation's harmonics.
As our example illustrates above, at least five and often more basis functions are required per simulation, with the
number increasing rapidly with mass as higher harmonics become more significant in band \cite{gwastro-spins-rangefit2010}.
However,  in principle no obstacle exists to constrained maximization over source orientation in an arbitrary reference
frame.

In practice, we expect that constrained
maximization of source orientation should be simplest in a frame aligned with the contours of source signal to noise ($\rho^2$)
and therefore in a frame roughly aligned with $\hat{V}$, the principal eigendirection of $\avL_M$.%
\footnote{For the purposes of this paper, we discuss constrained maximization of source orientation, assuming all
  harmonics are known from simulations or post-Newtonian calculations.  More broadly, by suitable interpolation across all masses and spins, a similar process of constrained
  maximization may allow waveform catalogs to reconstruct all the basis waveforms of precessing binaries.}
In this frame, the hermetian overlap matrix  $C_{AB}$ should generally have the simplest possible form
\begin{eqnarray}
C_{AB}^{-1} &\equiv& (A,B) 
\end{eqnarray}
where $A$ is shorthand for normalized basis functions of time, proportional to a subset of $ \WeylScalar_{lm}(t)$.  As a concrete example,
the eigenvectors of $C_{AB}$ using the $l=2$ basis states and our fiducial binary are fairly close to the basis states
themselves. 
 The reconstruction process for $\WeylScalar(\hat{n})$ involves  $C$ explicitly and
$C^{-1}$ implicitly:
\begin{eqnarray}
\WeylScalar(\hat{n}) &=& C_{AB}\rho_B(\hat{n}) \qmstate{B} \\
\rho_A &\equiv& (A,\WeylScalar)
\end{eqnarray}
The four free parameters in the signal enter through expressions involving $C^{-1}$.   First,  the parameters $\rho_A$  must reproduce the total signal $\rho^2 \equiv
(\WeylScalar,\WeylScalar)$ along this line of sight, or
\begin{eqnarray}
\rho^2 = \rho_A^* C_{AB}^{-1}\rho_B
\end{eqnarray}
The remaining three parameters correspond to specifying  the orientation of the source, with three Euler angles.  Two
of the Euler angles can be specified using the principal axes of $\avL_M$, which has the form
\begin{eqnarray}
\avL_M &=& \frac{C_{AB}^{-1}}{\rho^2}\left[\int d\Omega (\rho_A(\hat{n})^*{\cal L}_a {\cal L}_b \rho_B(\hat{n})) \right]
\end{eqnarray}
 The remaining Euler angle, the polarization angle,  is connected to the phase of $\rho$.
Though we adopt a multipolar basis in \emph{angle} to define the basis set in \emph{time}, the time domain basis
functions $\qmstate{A}$ are not generally orthogonal, so $\rho_A(\hat{n})$ are generally not proportional to spin-weighted
harmonics.\footnote{In special cases such as nonprecesing inspiralling binaries, the time-domain basis functions are
  orthogonal and $\rho_A$ are proportional to spin-weighted harmonics.  In this special case,  the integral on the right can be evaluated using representation theory of
SU(2) \cite{gwastro-mergers-nr-Alignment-ROS-Methods}.}
Conversely, only parameters in the corresponding 4-dimensional submanifold of the $2N$ (real) dimensional space are
consistent reconstructions of $\WeylScalar$ from this source.  

Therefore, the overlap matrix $C$ of the time-domain basis coefficients enters naturally into all reconstructions and constraints.
Their underlying algebra will be simplest if this $N\times N$ complex matrix takes block diagonal form.
However,  our arguments above suggest that no fixed reference frame, not even one aligned with $J$,  will make $C$ block diagonal at all masses.

\section{Conclusions}
Merging black hole binaries produce complicated orientation-dependent emission.  
For nonprecessing binaries, the orbital angular momentum defines a preferred orientation.  For these simple binaries, a generic signal
can be well-approximated by a linear combination of $\WeylScalar(\hat{L})$ and $\WeylScalar(-\hat{L})$, or just
one of the two functions $\WeylScalar(\pm \hat{L})$ away from the orbital plane. 
We demonstrate that these simplifying approximations break down for generic precessing binaries, by comparing  
different short, high-mass merger signals extracted along different orientations, using an extremely loose, data-analysis-motivated
discriminator.
Notably, our example shows that the final black hole spin angular momentum axis is emphatically \emph{not} the natural
emission orientation at late times.

For nonprecessing and some precessing binaries, most detection-weighted gravitational wave signal power ($\rho^2$) is radiated
along the total angular momentum axis.   Our counterexample also demonstrates that neither $J$, $L$, nor $S$ generally
corresponds to the direction of largest detectable gravitational wave signal.   By contrast, the peak of SNR ($\rho^2$)
versus orientation is nearly along the dominant eigendirection of  $\avL_M$ \cite{gwastro-mergers-nr-Alignment-ROS-Methods}.

For nonprecessing binaries, the orbit and kinematics define a preferred symmetry axis, relative to which the modulations
encoded in the waveform appear particularly simple.  For generic precessing binaries, particularly during the merger
phase, different physical concerns (kicks; precession) lead to different preferred orientations.

For data analysis purposes, we recommend an orientation along the peak of the network-weighted SNR along the line of
sight ($\rho^2$) or the nearly-identical location identified by the principal eigenvalue of $\avL_M$.
This frame will change with mass. 
However, we anticipate PTF-style constrained maximizations over source orientations will generally be simpler in this frame.
We will discuss the differences between this choice and the standard frame in a subsequent publication.

\ForInternalReference{

* No clear meaning to ``best orientation'' -- depends strongly on what you ask

* For data analysis purposes, interested in comparing wf against those from different orientations in the SAME
simulation

* In that sense, $\avL$ using \emph{both signs of frequency} is natural: it's the ability to distinguish orientations
when you use full information

* However, conceptually also useful to restrict to single sign
}

The breakdown of $J$ as a preferred orientation comes as no surprise.  As an extreme example, for sufficiently extreme-mass ratio binaries, the total angular
momentum is always dominated by the black hole spin, carrying no information whatsoever about the motion.  
In fact, the relevant preferred orientation depends on how the detector ``averages'' the orbit.  As a less extreme
example, for BH-NS binaries, the orbital angular momentum direction is the preferred direction on any sufficiently short
timescale.  For binaries whose masses and spins allow significant evolution in band, the relevant orientation will become (roughly)
a time-average of that vector \cite{gwastro-mergers-nr-Alignment-ROS-Methods}.  Only if the orbital angular momentum precesses
frequently about the total angular momentum \emph{in band} will the total angular momentum arise as a preferred orientation.
Even for high-mass-ratio binaries, precession can be irrelevant if it occurs out of band (e.g., before the detected
merger signal).
Finally, in this paper we introduced several concepts useful for interpreting the signal from precessing, unequal-mass
binaries.
First and foremost, we describe a complex overlap and its natural maximization in time and \emph{polarization}, a tool
which efficiently eliminates two extrinsic parameters of generic signals even in the presence of higher harmonics.
Second, we emphasize that $\psi_4$, being spin-weighted, naturally divides into two chiral parts;  this
decomposition  corresponds to positive and negative frequency in the fourier domain.

\appendix
\ForInternalReference{
\section{Direction diagnostics}

\subsection{Time domain}
* Peak of $|\psi_4|$ -- not necessarily useful

* $J$ initial and final

\subsection{Detection-weighted}

* More practical, reflect data analysis issues

* CHIRALITY (positive vs negative frequency)

(a) average orientation tensor described above

(b) $\psi_4$ dotted against other directions
}

\section{Chiral-weighted Orientation}
In general, the two extrema  of $\rho^2$ need not be antipodal.   Some methods for determining ``the'' optimal emission
direction treat $\hat{n}$ and $-\hat{n}$ symmetrically.   
To allow for different orientations of these two poles, we generalize the tensor $\avL_M$ introduced previously 
to treat contributions from $+$ and $-$ helicities separately, if desired.   For example, a matrix that sums up contributions from only right-handed emission is
\begin{eqnarray}
\label{eq:avL:Plus}
\left< L_{(a} L_{b)} \right>_{M,+} &=& 
 \frac{\int_0^{\infty} df \int d\Omega \FourierWeylScalar^*(f) L_{(a}L_{b)} \FourierWeylScalar(f)/S_h
  }{
   \int_0^\infty df \int d\Omega  |\FourierWeylScalar(f)|^2/S_h
}
\end{eqnarray}
 This construct correctly identifies the asymmetric extrema of $\rho_\pm$ seen in even nonprecessing binaries; see, e.g.,
the red dot in the bottom panel of Figure \ref{fig:Amplitudes:Frequency}. 

\section{Complex overlap and antipodal decomposition for symmetric quadrupole-dominated emission}

When one or a handful of modes dominates, the angular dependence of the overlap
$P(\hat{n},\hat{n}')$ can be approximated everywhere via a handful of global coefficients.
 As a practical example, if $\WeylScalar$ is  dominated by symmetric $l=|m|=2$ emission in the
 reference frame, the Weyl scalar can be expanded $\WeylScalar$ as
\begin{subequations}
\label{eq:OverlapVersusAngle:QuadrupoleDominant}
\begin{eqnarray}
\WeylScalar &=& \WeylScalar_{22}(t) \Y{-2}_{22}(\hat{n}) + \WeylScalar_{22}(t)^* \Y{-2}_{2-2}(\hat{n})  \\
P_{22} &\equiv& \left( \WeylScalar_{22}, \WeylScalar_{22} \right)/|\WeylScalar_{22}|^2 =1
\end{eqnarray}
where we have required the two antipodal terms to be complex conjugates, as for a nonprecessing signal.
In terms of these coefficients, we  find the overlap $P(\hat{n},\hat{n}')$  (before maximizing over time and phase) can be
well approximated by
\begin{eqnarray}
P &\simeq & |P_{22}| \frac{ \left[  Y_{2}^*Y_{2}'
 + 
  Y_{-2}^* Y_{-2}' \right]
}{
\sqrt{( |Y_{2}|^2 + |Y_{-2}|^2) ( |Y_{2}'|^2 + |Y_{-2}'|^2)}
 }  
\nonumber \\ &+& O( \left( \WeylScalar_{22},\WeylScalar_{22}^* \right) )\\
 &=& \frac{
  e^{2 i \phi}Y_{2}(\theta)Y_2 (\theta')
 + 
  e^{-2 i  \phi}Y_{-2}(\theta)Y_{-2}(\theta')
 }{
\sqrt{( |Y_{2}(\theta)|^2  + |Y_{-2}(\theta)|^2)(|Y_{2}(\theta')|^2+ |Y_{-2}(\theta')|^2)}
} \nonumber 
\end{eqnarray}
\end{subequations}
where we use the shorthand $Y_m\equiv \Y{-2}_{2m}(\hat{n})$ and similarly for $Y_m'$ to reduce superfluous subscripts
and where in the last line we factor out the common $e^{im\phi}$ from $\Y{-2}_{lm}$.
In the above expression we assume rapidly oscillating,  opposite-chirality terms like  $\left( \WeylScalar_{22},\WeylScalar_{22}^* \right)$
 cancel.\footnote{For a nonprecessing binary,  each sign of $m$ is nearly chiral,  having frequency content for the
   corresponding sign of $f$, and the complex conjugate of $-m$.  As a result, the $(2,2)$ and
  $(2,-2)$ modes are almost exactly orthogonal.}
We also make the trivial observation that the normalized overlap of a signal with itself is unity ($P_{22}=1$).
This purely geometrical expression is largest when the two vectors are coaligned with their respective reference frames
(i.e., when $\theta=\theta'=0$).
As an example, the special case of $\theta'=0$ has the form
\begin{eqnarray}
|P(\hat{z},\hat{n})| &=& \frac{\cos^4\theta/2}{\sqrt{\cos^8 \theta/2 + \sin^8 \theta/2}}
\end{eqnarray}

\begin{acknowledgements}
The authors have benefitted from feedback from Mark Hannam, Evan Ochsner, Diego Fazi, Andrew Lundgren,  Jolien Creighton, and Pablo Laguna.
DS is supported by NSF  awards PHY-0925345, PHY- 0941417, PHY-0903973 and TG-PHY060013N.
ROS is supported by NSF award PHY-0970074, the Bradley Program Fellowship, and the UWM Research
Growth Initiative.
\end{acknowledgements}

\bibliography{nr,ligo,gw-astronomy,gr-nr-vacuum,gr-nr-vacuum-kicks,gw-astronomy-mergers-nr,%
gw-astronomy-mergers-approximations,gw-astronomy-mergers,%
popsyn_gw-merger-rates,astrophysics-stellar-dynamics-theory,%
LIGO-publications,gw-astronomy-detection}
\end{document}